\documentclass[12pt,onecolumn,twoside]{opticajnl}
\journal{opticajournal} 

\setboolean{shortarticle}{true}

\usepackage{stackengine}
\usepackage{graphicx}

\usepackage{lineno}

\title{Subwavelength resolution using the near field of quantum emitters}

\author{Aziz Kolk\i ran$^{*}$}

\affil{Dept. of Engineering Sciences, \.{I}zmir Katip \c{C}elebi University,35620 \c{C}i\v{g}li, \.{I}zmir, Turkey}
\affil{aziz.kolkiran@ikcu.edu.tr}
\affil[*]{aziz.kolkiran@gmail.com}
\affil[*]{\href{https://orcid.org/0000-0002-7440-5428}{ORCID}}
\affil[*]{\href{https://sites.google.com/view/azizk/home}{web}}

\begin{abstract}
We propose a novel, to the best of our knowledge, approach to superresolution optical imaging by combining quantum optics and near-field optics. Our concept involves the utilization of single-photon quantum emitters to generate a stand-alone evanescent wave. We demonstrate that the quantum interference effects of single-photon emitters, in conjunction with their near-field, result in a higher resolution of subwavelength structures than systems that are only quantum enhanced or only near-field enhanced. We believe that nano-sized emitters could be employed to accomplish the goals of this research, taking into account the current progress in nanophotonics and quantum optics technology. 
\end{abstract}

\setboolean{displaycopyright}{false} 

\begin{document}

\maketitle


Quantum imaging utilizes a quantum system, such as a single photon or atom, as a probe to detect and measure the light emitted by an object. This allows for the production of highly detailed images with very high sensitivity and resolution. To increase the sensitivity and resolution of the imaging process, the quantum system is carefully manipulated using a variety of techniques, such as quantum entanglement and quantum interference. Numerous proposals have been made to improve different aspects of image formation, with only a few \cite{Boto,Thiel2007,Thiel2009,Moreau2019} improving the spatial resolution of the image, that is, the ability to image a physical object while surpassing the Rayleigh \cite{Rayleigh} or Abbe limit \cite{Abbe} of classical optics. This limit, a consequence of diffraction, is about half the wavelength. Recent experiments have yielded empirical support indicating that the attainment of subwavelength resolution in quantum imaging can be accomplished through the utilization of both correlated \cite{Moreau2014,Ndagano2022} and uncorrelated \cite{Israel2017,Lidke2005} photon sources. Many fluorophores are inherently single-photon emitters. This phenomenon is observable in multiple entities, including dye molecules and quantum dots. As an example, in the study conducted by Israel et al. \cite{Israel2017}, the authors utilized second-order correlation measurements in combination with an image localization algorithm to distinguish separate quantum dots that are not correlated. These methods rely on second- or higher-order correlations between field amplitudes to surpass classical limitations. Quantum interference can occur when a measurement cannot tell us which path a quantum state has taken, also known as "\emph{Which Path}" or "\emph{Welcher Weg}" information. For instance, if $N$ different detectors are used to detect $N$ photons from $N$ single-photon emitters in the far-field region, there are $N!$ ways that the $N$ photons can travel from the $N$ atoms to the $N$ detectors. The $N!$ quantum paths can differ in the optical phase, resulting in destructive or constructive interference between the $N$ photon amplitudes \cite{Thiel2007}. This can be used to acquire useful information about the spatial distribution of the source, even if it is smaller than the optical wavelength. The literature widely acknowledges that the quantum enhancement resulting from the indistinguishability of photon paths in spontaneously emitting systems is attributed to symmetric Dicke superpositions, referred to as superradiance \cite{Dicke54}. Superradiance involves the coordinated action of emitters, leading to the emission of radiation that is more intense, focused, and coherent compared to radiation emitted individually by emitters, and can occur with or without direct interactions among emitters. This phenomenon is currently of significant interest and has been effectively demonstrated in various quantum emitters experimentally \cite{Koong22,Solano17,Trebbia22}. This study aims to illustrate how these characteristics of closely positioned systems of two and four emitters can be utilized to achieve quantum superresolution.

In contrast to the Rayleigh criterion, near-field imaging has been shown to improve resolution \cite{Vigoureux92_1, Wolf70, Agarwal_98}. It is well known that a small object diffracts both propagating and evanescent modes of high spatial frequency. Near-field optical methods can convert these high-spatial-frequency evanescent waves into radiating waves, thus allowing for improved resolution. This enables the extraction of details about an object's microscopic structure, which may be a small fraction of the source wavelength. Near-field detection methods \cite{Betzig91,Oshikane2007} require the use of very small detectors placed a distance from the object in the subwavelength range. On the other hand, the reciprocity theorem can be applied to a small object placed in an evanescent field, which converts part of this field into propagating waves \cite{Vigoureux92_2}. This allows subwavelength structures to be imaged in the far zone without the need for detectors to be placed in the near zone. In reference \cite{Lai16}, it was demonstrated that high-refractive-index microspheres can be used to form superresolution images in microscopy by collecting high spatial-frequency signals at near-field distances. In a previous study by the authors \cite{kolkiran2012}, a single-photon emitter was used as an independent evanescent wave source to achieve this effect.
\begin{figure}[t]
\centering 
\includegraphics[width=0.65\linewidth]{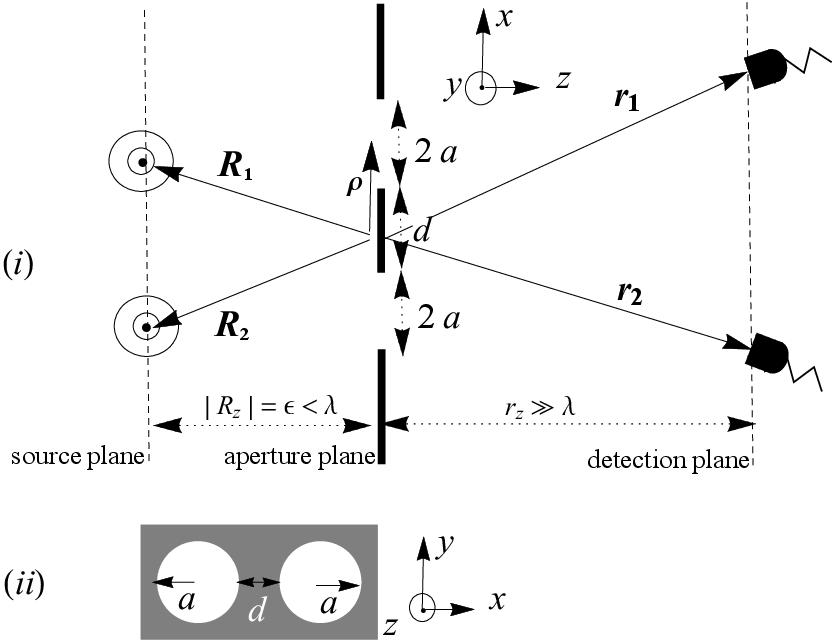}
\caption{($i$) A crosssection of the imaging setup with coordinates. 
($ii$) Displays the apertures in the $xy$ plane. Throughout the study, aperture radii $a=\lambda/2$ and aperture distance $d=\lambda/4$. }
\label{fig1}
\end{figure}


Inspired by these studies, we propose to combine quantum enhancement with near-field effects to achieve a kind of multiplier effect for super-resolution. We integrate the methods described in Refs. \cite{Israel2017, Lai16} into a unified framework. Furthermore, we propose implementing the coincidence detection scheme to exploit the benefits of higher-order quantum interference phenomena. To initiate the analysis, we will examine two individual sources of single photons that are located at distances smaller than the wavelength from an array of apertures (see Fig. \ref{fig1}) The resolution will be further improved by the quantum interference effects of the two emitters in addition to their near-fields. We follow the basic models in Refs. \cite{Thiel2009,kolkiran2012}. The sources in the proposed setup can be considered as two atoms with identical energy levels ($|g\rangle$ and $|e\rangle$). These atoms, located at $\mathbf{R_1}$ and $\mathbf{R_2}$, can be excited by a laser pulse and emit photons spontaneously. Photons are detected by detectors located at positions $\mathbf{r_1}$ and $\mathbf{r_2}$ in the far-field region. An aperture array is placed between the atoms and the detectors for imaging. This aperture array consists of two identical circular apertures placed very close to each other. Two photons emitted from each atom are diffracted by the apertures and are recorded by two detectors in the far-field region. For simplicity, the planes that contain the atoms, the object, and the detectors are parallel. We will also fix the detectors and the source in the $xz$ plane to calculate the diffraction patterns. This means that the $z$ components of all $\mathbf{r_i}$'s and $\mathbf{R_j}$'s are the same, that is \(|r_{1z}|=|r_{2z}|=:r_{z}\) and \(|R_{1z}|=|R_{2z}|=:R_{z}\equiv\epsilon\). We can measure the joint probability of finding one photon at $\mathbf{r_1}$ and another at $\mathbf{r_2}$ by calculating the second-order correlation function.
\begin{equation}\label{G2}
G^{(2)}({\bf{r_1,r_2}}) = \langle E^{(-)}(\mathbf{r_1})E^{(-)}(\mathbf{r_2})E^{(+)}(\mathbf{r_2})E^{(+)}(\mathbf{r_1})\rangle,
\end{equation}
where $E^{(+)}(\mathbf{r_i})=[E^{(-)}(\mathbf{r_i})]^{\dag}$, denotes the
positive frequency part of the electric field amplitude at point
$\mathbf{r_i}$. Note that our scheme does not require coincident detection. The requirement is that all photons emitted by the 2 atoms are detected by the 2 detectors. The exact time of detection does not affect the contrast or the resolution of the correlation signal. 

\begin{figure}[ht]
 \centering
 \includegraphics[width=0.8\linewidth]{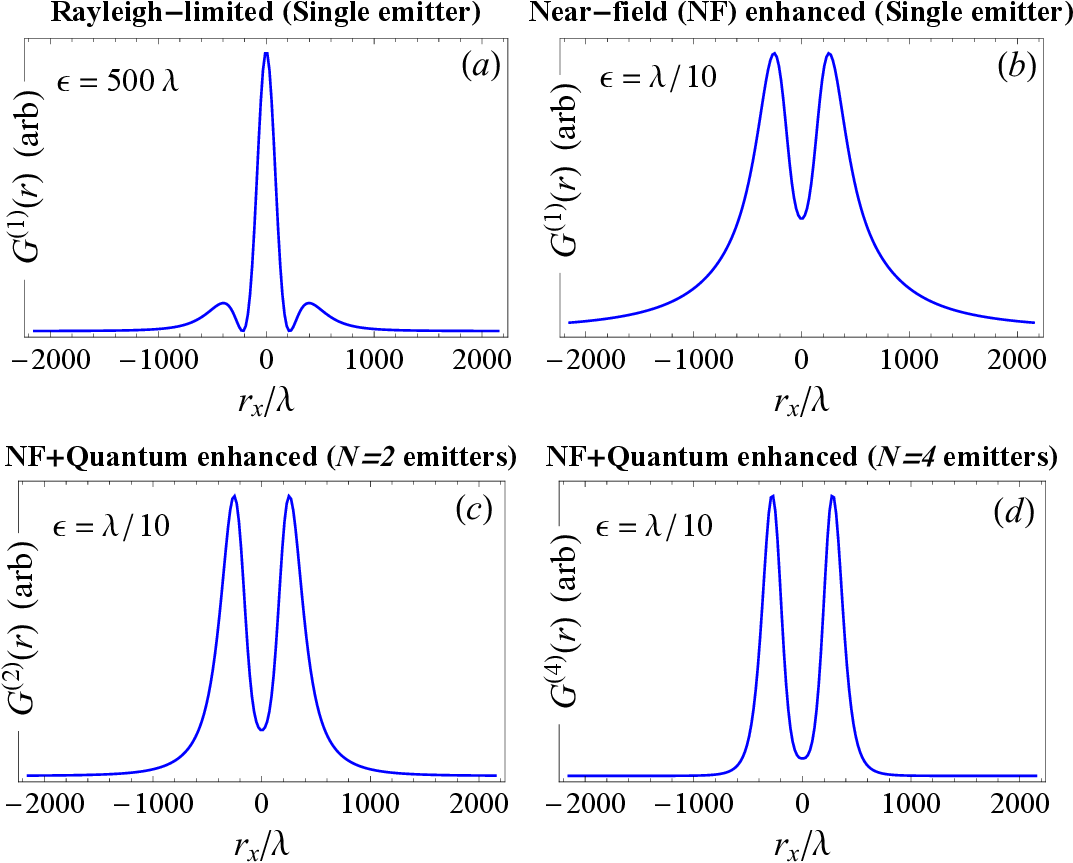}
  \setlength{\unitlength}{0.008\linewidth}
    \begin{picture}(100,0)    
        \put(35,53) 
        {\includegraphics[width=12\unitlength]{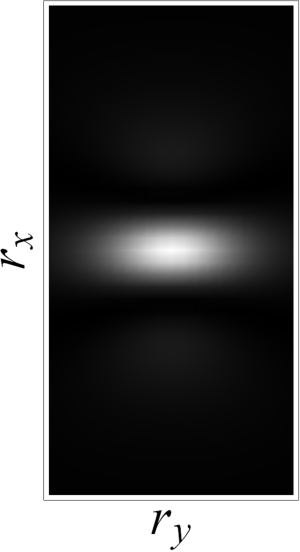}}
        \put(85,53)  
        {\includegraphics[width=12\unitlength]{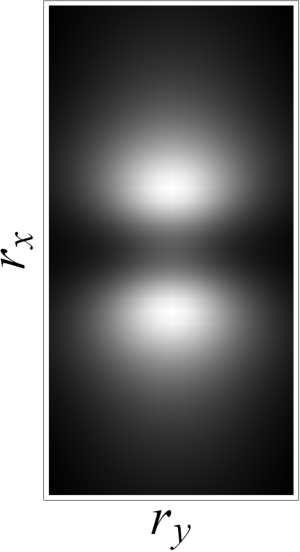}}
        \put(35,11) 
        {\includegraphics[width=12\unitlength]{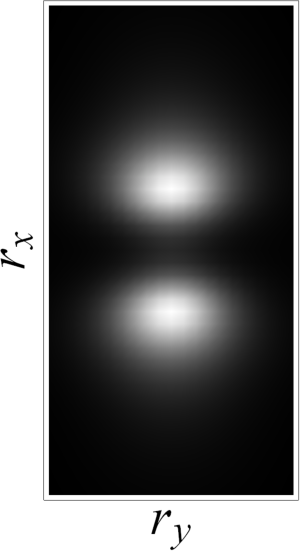}}
        \put(85,11) 
        {\includegraphics[width=12\unitlength]{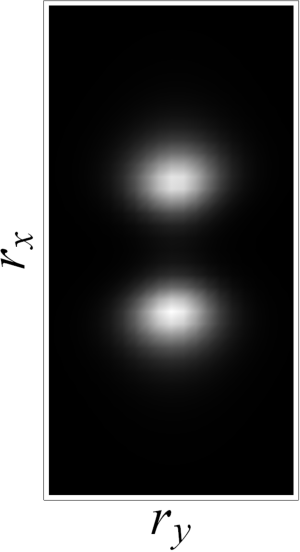}}
    \end{picture}
 \caption{Diffraction patterns of two sub-wavelength-sized apertures. The diffraction patterns were calculated using first-order (($a$) and ($b$)), second-order, ($c$), and fourth-order, ($d$), correlation functions. The detectors move along the $x$ axis and $r_z=500\lambda$ in all figures. In ($a$), the source plane and the detection plane are in the far zone. In ($b$), ($c$) and ($d$) the source plane is located at $\lambda/10$. 
 The insets illustrate the images in two dimensions.}
 \label{fig2}
\end{figure}

Next, we calculate the field amplitudes generated by the point sources and diffracted by the apertures. Since radiation from a localized source can be considered as a diverging spherical wave, we can assume that photons from the atoms are guided by $\exp(ikr)/r$ with a suppressed time dependence $\exp(-i\omega t)$,
where $k=\omega/c=2\pi/\lambda$. Using scalar diffraction theory, we can now calculate the diffraction of the electric field by the apertures. Assuming that the apertures are inside a medium that is linear, isotropic, homogenous, nondispersive and nonmagnetic, all components of the electromagnetic field are uncoupled and the space-dependent part $U(\mathbf{r})$ satisfies the Helmholtz wave equation $(\nabla^2+k^2)U(\mathbf{r})=0$ at each source-free point. For diffraction by planar screens, the equation can be solved by following Kirchhoff in applying
Green's theorem.
     \begin{align}\label{Kirchhoff1}
 U(\mathbf{r_i},\mathbf{R_j})=-\frac{1}{4\pi}\iint_S dS\left\{\frac{\partial U}{\partial n}
 \left[\frac{e^{ik|\mathbf{r_i}-\boldsymbol{\rho}|}}{|\mathbf{r_i}-\boldsymbol{\rho}|}\right]
 -U\frac{\partial}{\partial n}\left[\frac{e^{ik|\mathbf{r_i}-\boldsymbol{\rho}|}}{|\mathbf{r_i}-\boldsymbol{\rho}|}\right]\right\},
 \end{align}
where the integral is taken over the plane of the aperture and
$\partial/\partial n$ signifies a partial derivative in the
outward normal direction at each point on $S$. The vectors $\mathbf{r_i}$ and $\mathbf{R_j}$ extend from the center of the apertures to the $i$'th detector and $j$'th
source, respectively, and $\boldsymbol{\rho}$ scans the
aperture plane (see Fig. \ref{fig1}). The field amplitude $U$ in
the aperture plane $S$ of the $j$' th
source is given by the spherical wave
$\exp(ik|\mathbf{R_j}-\boldsymbol{\rho}|)/{|\mathbf{R_j}-\boldsymbol{\rho}|}$. If the source and the detection plane are both located in the far zone, where $ |\mathbf{r}-\boldsymbol{\rho}| , |\mathbf{R}-\boldsymbol{\rho}| \gg k \boldsymbol{\rho} ^2/2$, then the intensity, or the first-order correlation function, $G^{(1)}({\bf{r}})=\langle E^{(-)}(\mathbf{r})E^{(+)}(\mathbf{r})\rangle= |U(\mathbf{r})|^2$, can be calculated, resulting in the Rayleigh criterion for object resolution. This means that two apertures separated by a distance smaller than the wavelength cannot be resolved. This is shown in the imaging of closely spaced apertures in Fig. \ref{fig2}(a). 

Next, if we set the position of the source plane having a sub-wavelength distance to the aperture plane, the $|\mathbf{R_i}-\boldsymbol{\rho}|\gg k|\boldsymbol{\rho}|^2/2$ approximation is no longer valid and the Eq. (\ref{Kirchhoff1}) becomes
   \begin{align}\label{Kirchhoff2}
U(\mathbf{r_i},\mathbf{R_j})=-\frac{\Phi}{r_{i,z}\lambda}\iint_S d\rho_xd\rho_ye^{i\frac {k}{r_{i,z}}\left (\rho_xr_{i,x} + \rho_yr_{i,y} \right)}
 \frac{e^{ik|\mathbf{R_j}-\boldsymbol{\rho}|}}{|\mathbf{R_j}-\boldsymbol{\rho}|}\frac{R_{j,z}}{|\mathbf{R_j}-\boldsymbol{\rho}|}\left(1-\frac{1}{ik|\mathbf{R_j}-\boldsymbol{\rho}|}\right),
 \end{align}
where $\Phi=\exp\{ikr_{i,z}+ik(r_{i,x}^2+r_{i,y}^2)/2r_{i,z}\}$ is a phase factor and for only one source, the indices $i$'s and $j$'s are to be removed. The integral given in Eq. (\ref{Kirchhoff2}) cannot be solved for an analytical expression, but one can easily see that it contains an infinite spectrum of spatial frequencies $k_x$ and $k_y$ satisfying $k_x^2+k_y^2>k^2$. To analyze the angular spectrum of the field, one can expand the spherical wave contained in the integral into an angular spectrum of plane waves, as given by Weyl's integral \cite{Weyl},
\begin{equation}\label{weyl decomposition}
\frac{e^{ik|\mathbf{R}-\boldsymbol{\rho}|}}{|\mathbf{R}-\boldsymbol{\rho}|}\!\!=\!\!\frac{i}{2\pi}\!\!\iint_{-\infty}^{+\infty}
\!\!\frac{dk_xdk_y}{k_z}e^{-i\left[k_x(R_x-\rho_x)+k_y(R_y-\rho_y)+k_zR_z\right]},
\end{equation}
 where
\begin{equation}
k_z = \left\{
\begin{array}{rl}\label{kz}
\sqrt{k^2-k_{||}^2} & \text{for } k_{||}^2\leq k^2,\\
i\sqrt{k_{||}^2-k^2} & \text{for } k_{||}^2> k^2,
\end{array} \right.
\end{equation}
$R_z=-\epsilon$ is the source-aperture distance on the $z$-axis and $k_{||}^2=k_x^2+k_y^2$. The phase factor $\Phi$ given in Eq. (\ref{Kirchhoff2}) ensures that the higher spatial frequency evanescent components ($k_{||}^2> k^2$) contribute to the signal in the far zone, leading to superresolution. This can be explained as follows: From the usual theory of diffraction, a small object illuminated with a propagating wave generates diffracted evanescent modes (in which $k_{||}^2> k^2$), and conversely, by applying the reciprocity theorem, a small object located in an evanescent field (or a near field) converts part of this field into propagating waves \cite{Guerra87, Guerra90,Vigoureux92_1}. An aperture located in the vicinity of a point source diffracts the evanescent part into radiation. Thus, the subwavelength details of the aperture are to be
carried to the far zone and the method does not require point-by-point, time-consuming scanning detection. The result of this method was shown by the previous work of the authors \cite{kolkiran2012}. Fig. \ref{fig2}(b) is the result of numerically solving the absolute square (intensity) of Eq. (\ref{Kirchhoff2}). It demonstrates that a detector located in the far zone can distinguish between two closely spaced sub-wavelength-sized circular apertures. 

Now consider two atoms that are located at a distance of less than one wavelength behind the plane of the aperture. For imaging, we now use two detectors with a coincidence detection scheme. This is calculated using the second-order correlation function given by Eq. (\ref{G2}). Detecting two photons from two single-photon sources would lead to a typical \emph{"which-path"} interference in quantum optics. It is well-known that higher-order correlations in quantum imaging always lead to an increase in resolution. In addition to the quantum interference effects, it is natural to expect a further increase in resolution due to the near-fields. In Fig. \ref{fig1}, two atoms contribute to the electric field at $\mathbf{r_i}$, each giving rise to a field $U(\mathbf{r_i},\mathbf{R_j})$ of the form given in Eq. (\ref{Kirchhoff2}). We can thus write the total positive frequency part of the field contributing to the correlation signal at $\mathbf{r_i}$ produced by two point sources at $\mathbf{R_1}$ and $\mathbf{R_2}$ as  
\begin{equation}\label{electric_field_operator}
    E^{(+)}(\mathbf{r_i})=\frac{1}{\sqrt{2}}\left\{U(\mathbf{r_i},\mathbf{R_1})|g\rangle_1\langle e|+U(\mathbf{r_i},\mathbf{R_2})|g\rangle_2\langle e| \right\},
\end{equation}
where the atomic operator $|g\rangle_j\langle e|$ describes the de-excitation of the $j$'th atom. With the initially excited state of the atoms $|e_1,e_2\rangle$ and Eq. (\ref{electric_field_operator}), the second order correlation function of Eq. (\ref{G2}) can be written as follows:
  \begin{align}\label{G2 explicit}
G^{(2)}({\bf{r_1,r_2}})&=\langle e_1,e_2|E^{(-)}(\mathbf{r_1})E^{(-)}(\mathbf{r_2})E^{(+)}(\mathbf{r_2})E^{(+)}(\mathbf{r_1})|e_1,e_2\rangle\nonumber \\
&=1/4|U(\mathbf{r_1},\mathbf{R_1})U(\mathbf{r_2},\mathbf{R_2})+U(\mathbf{r_1},\mathbf{R_2})U(\mathbf{r_2},\mathbf{R_1})|^2.
 \end{align}
In addition to the super-resolution power of the near-field inherent in the amplitudes $U(\mathbf{r_i},\mathbf{R_j})$, our goal now is to create a quantum interference effect between the two-photon paths so that we can obtain a faster optical phase change in the signal, leading to higher-resolution imaging. In general, for $N$ photon detection, modulation in the optical signal has been shown to increase $N$ times relative to the classical case \cite{Thiel2007}. The signal given in Eq. (\ref{G2 explicit}) can only be calculated using numerical methods, since the field amplitudes $U(\mathbf{r_i},\mathbf{R_j})$ have no closed-form expressions.  Since $G^{(2)}({\bf{r_1,r_2}})$ can be calculated analytically for the case where the source and detectors are in the far zone, in the reference \cite{Thiel2009} it was shown that a factor of $2$ can be achieved in modulating the phase change if the positions of the two atoms were chosen as $R_{1x}=0, R_{2x}= \frac{\pi R_z}{ka}$. Here, $a$ denotes the aperture dimension in the $x$ direction. In our setup, to ensure that the two photons are indistinguishable, the two atoms are positioned so that $R_{1x}=0, R_{2x}=\frac{\pi R_z}{k(4a+d)}$. More information on how to estimate the relative positions is available in Sections 1-2 of \href{supp}{supplementary document}. To double the rate of phase change, the detector locations should be $r_{2x}=-r_{1x}$. There are now more alternatives for two photons to reach two detectors. The indistinguishability of these alternatives in detector measurements has led to a quantum advantage over the single-source situation, resulting in a marked enhancement in resolution. Fig. \ref{fig2}(c) displays the $G^{(2)}({\bf{r_1,r_2}})$ signal in two ways: along the $x$ axis and as a two-dimensional image on the $xy$ plane. The 2D image is obtained by capturing photon pairs using photon detectors located at $r_{1x}=r_x$ and $r_{2x}=-r_x$ at every line of constant $r_y$. 
The enhancement in resolution comes clearly from the quantum character generated by the measurement process after the detection of the first photon. It has been shown in \cite{Dur,Skornia} that
just before the detection of the second photon, the atomic
system is in a 2-particle $W$ state with one excitation. This is also true in a system of $N$ atoms with $N$ detectors. The result shown in Fig. \ref{fig2}(c) proves that it is possible to exploit both near-field (evanescent electromagnetic field) and quantum interference effects simultaneously to further improve the optical resolution compared to the classical case. The third section of  \href{supp}{supplementary document} provides a detailed explanation of how the enhancement occurs in the two-photon case, specifically concerning the indistinguishability of the two-photon paths.

\begin{figure}[t]
\centering 
\includegraphics[width=0.8\linewidth]{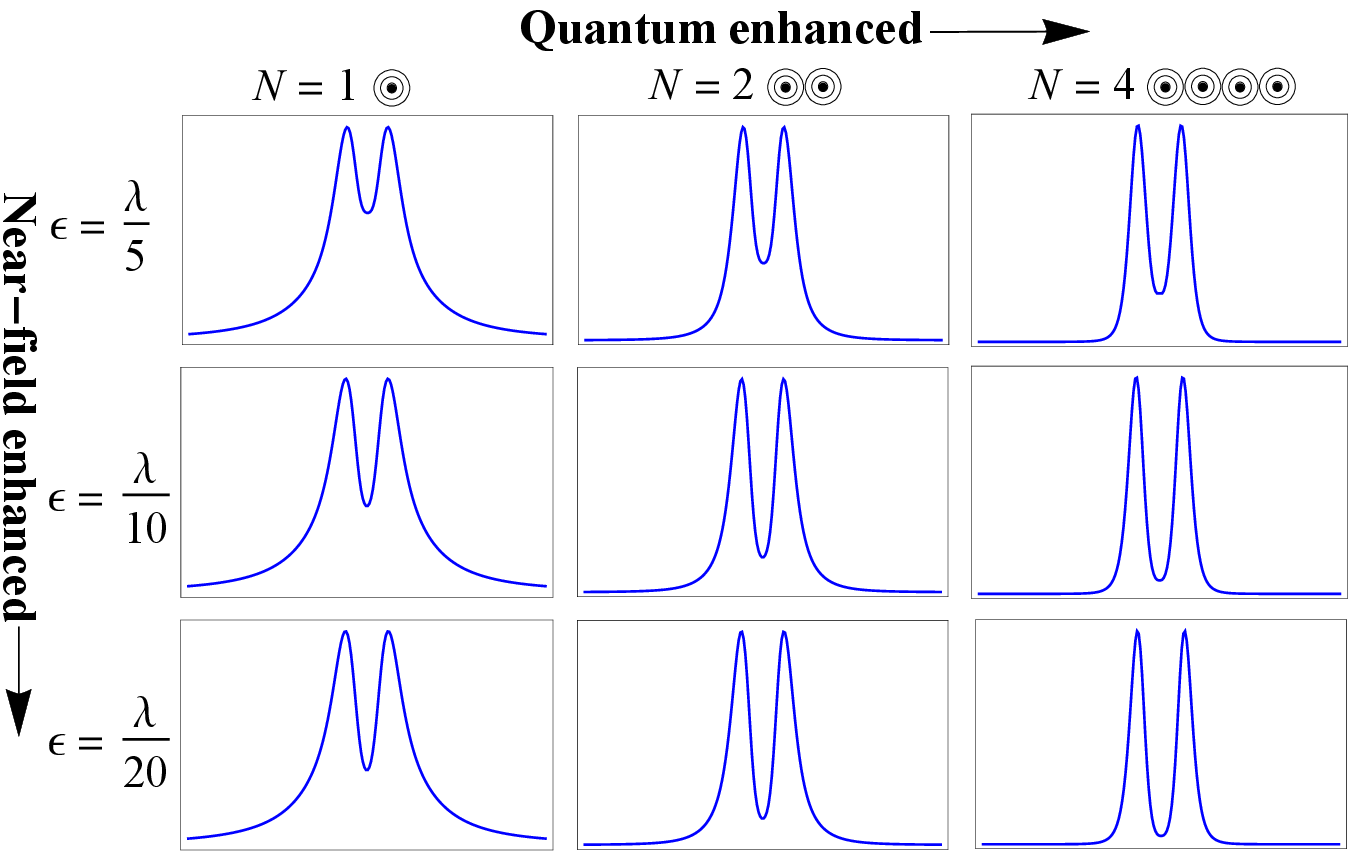}
\caption{This matrix displays the resolution pattern of hybrid super-resolution depending on the number of co-emitters and the co-source-aperture distance. The degree of quantum correlation increases from left to right, whereas the near-field effect increases from top to bottom. 
All graphs have $r_z=500\lambda$.}
\label{fig3}
\end{figure}

Now, let us increase the number of emitters and detectors to $N=4$. In this case, the $4th$ order correlation signal, $G^{(4)}(\bf{r_1,r_2,r_3,r_4})$, will be the signal to be measured for the photons of the $4$ emitters. We choose the position of the emitters as $ R_{1x}=-\frac{\pi R_z}{k(4a+d)}, \; R_{2x}=0, \; R_{3x}=\frac{\pi R_z}{2k(4a+d)}, \; R_{4x}=\frac{\pi R_z}{k(4a+d)}$ and let the detector positions be $r_{1x}=-r_{2x},\; r_{3x}=-r_{1x}+\frac{\pi r_z}{k(4a+d)}$ and $r_{4x}=r_{1x}+\frac{\pi r_z}{2k(4a+d)}$. Again, these parameters are chosen in this way to increase the phase modulation owing to quantum interference effects compared to classical phase modulation. The four photons from each source will reach the four detectors through a total of twenty-four ($4!=24$) distinct pathways. For all emitters prepared in the excited state $|e\rangle$, we get,
\begin{equation}\label{G4}
G^{(4)}({\bf{r_1,r_2,r_3,r_4}})=A|\xi ({\bf{r_1,r_2,r_3,r_4}})|^2,
\end{equation}
where
\begin{equation}\label{xi}
\xi ({\bf{r_1,r_2,r_3,r_4}})=\sum^4_{\substack{a_1\ldots a_4=1\\a_1\neq\ldots\neq a_4} }\prod^4_{\mu=1}U({\mathbf{r}}_{a_{\mu}},\mathbf{R}_{\mu}),
\end{equation}
and $A$ is a constant. Fig. \ref{fig2}(d) illustrates the difference in resolution when there are four emitters compared to two emitters. As the number of emitters increases, the resolution increases significantly, allowing the images of the apertures to be seen with perfect clarity. Fig. \ref{fig3} collectively shows the variation of resolutions with the number of emitters and the emitter-aperture distance. This comparatively demonstrates near-field and quantum interference effects in a single setup. Superresolution can be achieved by using both near-field and quantum correlations at the same time. The calculations assume that the emitters are of point size, which means that, theoretically, the epsilon ($\epsilon$) value can be as small as possible. However, further investigations are necessary using rigorous wave equations to establish the mathematical relationship between our results and the theoretical resolution limit. As the number of emitters and detectors increases, higher-order correlations lead to higher resolution. However, as $N$ increases, the system becomes more challenging to control and measure experimentally. In the four-emitter configuration, it is possible to decrease the resolution to the same level as in the two-emitter and single-emitter arrangements by altering the correlation measurements of the four detectors. This is thoroughly discussed in the fourth section of \href{supp}{supplementary document}.

In fact, there is a study \cite{Ben-Aryeh_2003} that makes a comparison between near-field super-resolution optical effects and effects obtained by non-classical electromagnetic states. Returning to the angular decomposition of the near field of the dipole emitter given by Eqs. (\ref{weyl decomposition}) and (\ref{kz}), we see that the general solution in the aperture plane can involve a superposition of modes propagating in both the $+x$ and $-x$ directions (these are demonstrated by the coordinates $\rho_x$ and $-\rho_x$ in the aperture plane). For example, consider the component where we are confined to the $y=0$ axis and $k_x=5k$. In this case, to fulfill the condition $k_x^2+k_y^2+k_z^2=k^2$, $k_z$ will be imaginary with a value of $i4.9k$ and there will be a superposition in the field amplitude as follows:
\begin{equation}
 U(r,R)\propto \left[exp(-i5k\rho_x)+exp(i5k\rho_x)\right] exp(-4.9k_zR_z).
\end{equation}
This superposition describes a standing wave with spatial modulation $\lambda/5$. If the emitters are chosen close enough to the aperture plane, which in our setup they are, and $R_z<\lambda$, the measurable magnitude of this superposition will increase. Therefore, considering only this component, the resolution power will increase by a factor of 5 in the x-direction compared to conventional imaging. We think that our result can shed some light on this matter.

To summarize, this study has illustrated that the combination of near-field effects and quantum correlations can be employed to achieve super-resolution in a single configuration. The super-resolution is quantified using detectors positioned in the far field, thereby eliminating the requirement for scanning sub-wavelength-sized tip detectors. By augmenting the number of independent emitters and detectors capable of correlated measurements, we have demonstrated the possibility of achieving both quantum super-resolution and near-field super-resolution within the same setup.This hybrid superimaging technique will also provide insight into the nonclassical properties of near-field electromagnetic components. The utilization of high-refractive-index structures \cite{Lai16} in conjunction with quantum dots that emit single photons, along with the measurement of photon-photon correlations \cite{Israel2017}, makes them inherent contenders for the implementation of this proposal. In summary, we have demonstrated that near-field and quantum super-resolution can be achieved in a single setup without the use of scanning-type detectors or entangled photons.
\paragraph{Funding}
A. K. thanks the Turkey Scientific and Technological Research Council (TÜBİTAK) grant no. 110T321 for supporting this research.



\bibliographyfullrefs{main}
\pagebreak
\section*{supplementary document}

This document provides supplementary material for \emph{ Subwavelength resolution using the near-field of quantum emitters.}

\section{The estimation of the relative position of the emitters}

\begin{figure}[htbp]
\centering
\fbox{\includegraphics[width=0.8\linewidth]{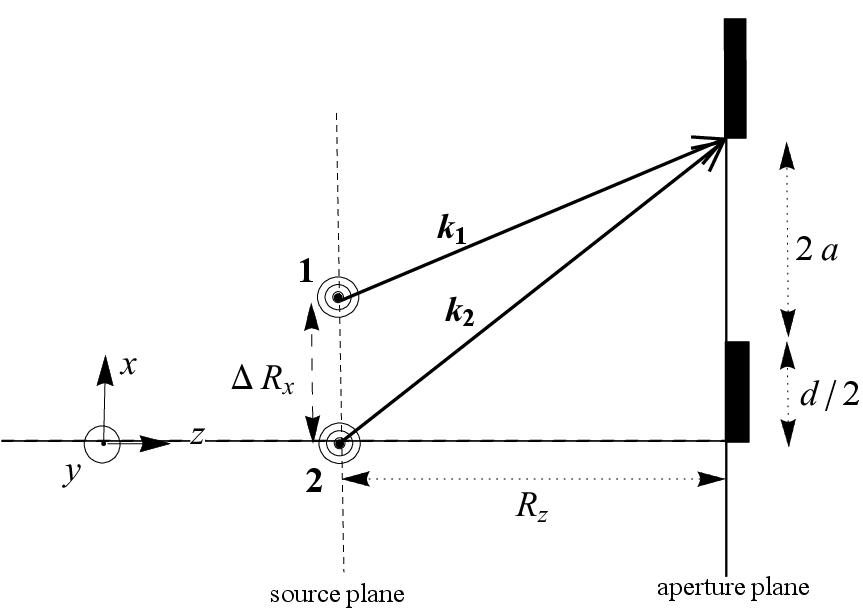}}
\caption{The figure shows the $\mathbf{k}$ vectors and the distances for the estimation of the relative position of the emitters for $N=2$ .}
\label{sfig1}
\end{figure}

To take advantage of the indistinguishability of photon paths, which offers a quantum advantage when multiple sources are employed, the sources must be situated close to one another. The indistinguishability of photons can be established using the uncertainty principle. The position uncertainty of each photon in the xy plane (here we consider the x-axis for simplicity) should overlap with the other. Fig. \ref{sfig1} illustrates the vectors and distances necessary for the uncertainty calculations of the two photons in the part of the setup above the $x$-axis. 

To estimate the value of $R_x$, we use the vectors and their components in Fig. \ref{sfig1}. In the figure $\mathbf{k_1}$ and $\mathbf{k_2}$ represent the wave vector of the sources and $|\mathbf{k_1}|= |\mathbf{k_2}|=k=2\pi/\lambda$. For any of the vectors $\mathbf{k}$,
\begin{equation}\label{kxkzratio}
     \frac{|k_x|}{|k_z|}\approx\frac{2a+d/2}{R_z}
\end{equation}
ratio is acceptable for typical values. In particular, this ratio requires $|k_x|>|k_y|$ for decreasing values of $R_z$, which is expected for the near field. We note that in the setup the $z$ component of the near-field wave vector of the sources will be imaginary. In this case, for example, in the $xz$-plane 
\begin{equation}\label{krelation}
     k_x^2-k_z^2=k^2=(2\pi/\lambda)^2,
\end{equation}
can be written. As sources approach the aperture, $|k_x|=p\cdot k$ and $|k_z|=(\sqrt{p^2-1})k$ for any $p>1$, that is, $k_x$ will exceed $k=2\pi/\lambda$. If we consider the $k_x$ component as the largest uncertainty of the photon in the $x$ direction, 
\begin{equation}\label{uncertainty}
      \Delta R_x\cdot k_x\geq\frac{\pi}{2}.
\end{equation}
Therefore, from $R_x=\pi/(2k_x)$ together with Eq.(\ref{kxkzratio}) , we get
\begin{equation}
       \Delta R_x=\frac{1}{\sqrt{p^2-1}}\frac{\lambda R_z}{2(4a+d)}.
\end{equation}
Therefore, we can take $\delta=\frac{\lambda R_z}{2(4a+d)}$ as the characteristic value of the uncertainty of the position of the photon in the $x-$direction. When two photons are placed inside this uncertainty, indistinguishability is achieved. 

\section{Hybrid resolution patterns versus interatomic distance}

\begin{figure}[htp]
\centering
\fbox{\includegraphics[width=0.8\linewidth]{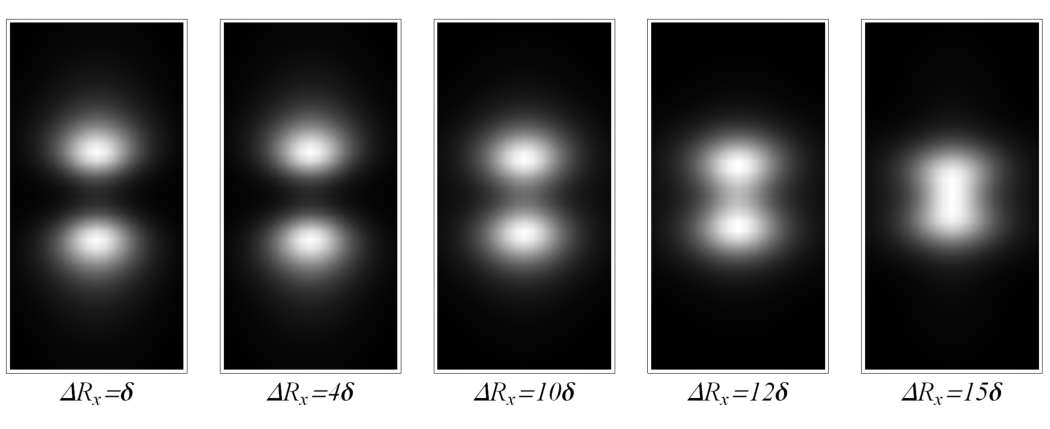}}
\caption{The changes in the two-dimensional resolution patterns of two emitters when the distance between them is altered are examined. The most optimal super-resolution is achieved when the distance between sources is $\delta$ or less. As the gap between them grows, the clarity diminishes and vanishes completely at around $\Delta R_x=15\delta$. The detector positions along the $y$-axis is the horizontal one and the detector positions along the $x$-axis is the vertical one in all patterns. The range of the $x$ axis is from $-1000\lambda$  to $+1000\lambda$, while the range of the $y$ axis is from $-500\lambda$ to $+500\lambda$. The distance between the source and the aperture planes is set to $\lambda/10$ while the detectors are at $r_z=500\lambda$.
}
\label{sfig2}
\end{figure}

The experimenter may be concerned with how the relative positioning of the sources affects the resolution when calculating second-order correlation signals.  The optimal inter-source distance is determined by multiplying the ratio of the distance between the source plane and the aperture plane to the size of the aperture by one-half of the wavelength, as calculated in the preceding section. Imaging calculations of the two-atom source ($N=2$) yielded the best resolution when the ratio $\delta=\frac{\pi R_Z}{k(4a+d)}$ was used. If the distance between the sources is larger than the ratio, the resolution can be expected to decrease as the indistinguishability of the photon paths is impacted. We illustrate how the plot of the signal $G^2{(\mathbf{r_1,r_2})}$ varies with the distance between the sources in Fig. \ref{sfig2}.

Imaging calculations of diatomic sources ($N=2$) give the best resolution when the distance between sources $\Delta R_x$ is less than or equal to $\delta$.  As $\Delta R_x$ increases to values ten times $\delta$, the resolution gradually decreases, and when it grows to about 15 times, the resolution completely disappears. 

\section{Two-photon pathways}

\begin{figure}[htb]
\centering 
\includegraphics[width=0.8\linewidth]{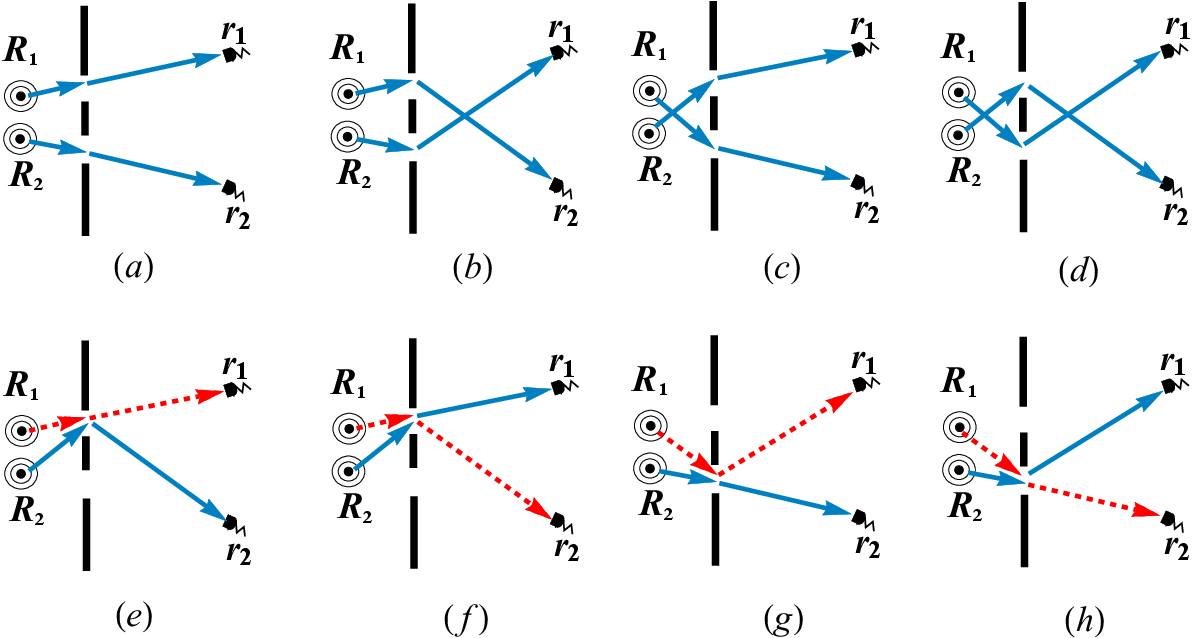}
\caption{All possible paths for detecting the two photons for imaging two apertures.  In ($a$)-($d$) the two photons pass through separate apertures, while in ($e$)-($h$) they both pass through the same aperture.}
\label{sfig3}
\end{figure}

We can interpret the enhancement that comes from the two-photon case in the following way: Two photons are sent from two atoms to two detectors in two distinct situations. The first occurs when both photons pass through the apertures separately (see Fig.\ref{sfig3}(a)-(d)), and the second occurs when both pass through the apertures together (see Fig.\ref{sfig3}(e)-(h)). The particular positioning of the atoms relative to each other suggests that the first case is subject to destructive quantum interference, while the second case is subject to constructive interference. Physically, this can be interpreted as the photon sources selectively illuminating only one aperture at a time, while the other remains dark, as can be clearly seen in Fig. \ref{sfig3}(e)-(h). This is reminiscent of the superposition $NOON$ with $N=2$ in quantum interferometers \cite{Boto}.

\section{Changing the detector correlations}

One of the conditions for achieving super-resolution is the use of multidetector correlations. In correlated measurements, the optical phase modulation rate is maximized by scanning the detectors at specific positions \cite{Thiel2007}. In this section, we will attempt to elaborate on this relationship by manipulating the detector correlations that determine the optical phase modulation rate at multiple sources. 
Our objective is simply to show superresolution patterns of fourth-, second-, and first-order field correlations in a four-atom system by altering the detector measurement approach. Thus, with sources placed in the near field, quantum-enhanced superresolution of any order can be demonstrated in a single setup by simply changing the measurement method.  
\begin{figure}[t]
\centering
\includegraphics[width=0.8\linewidth]{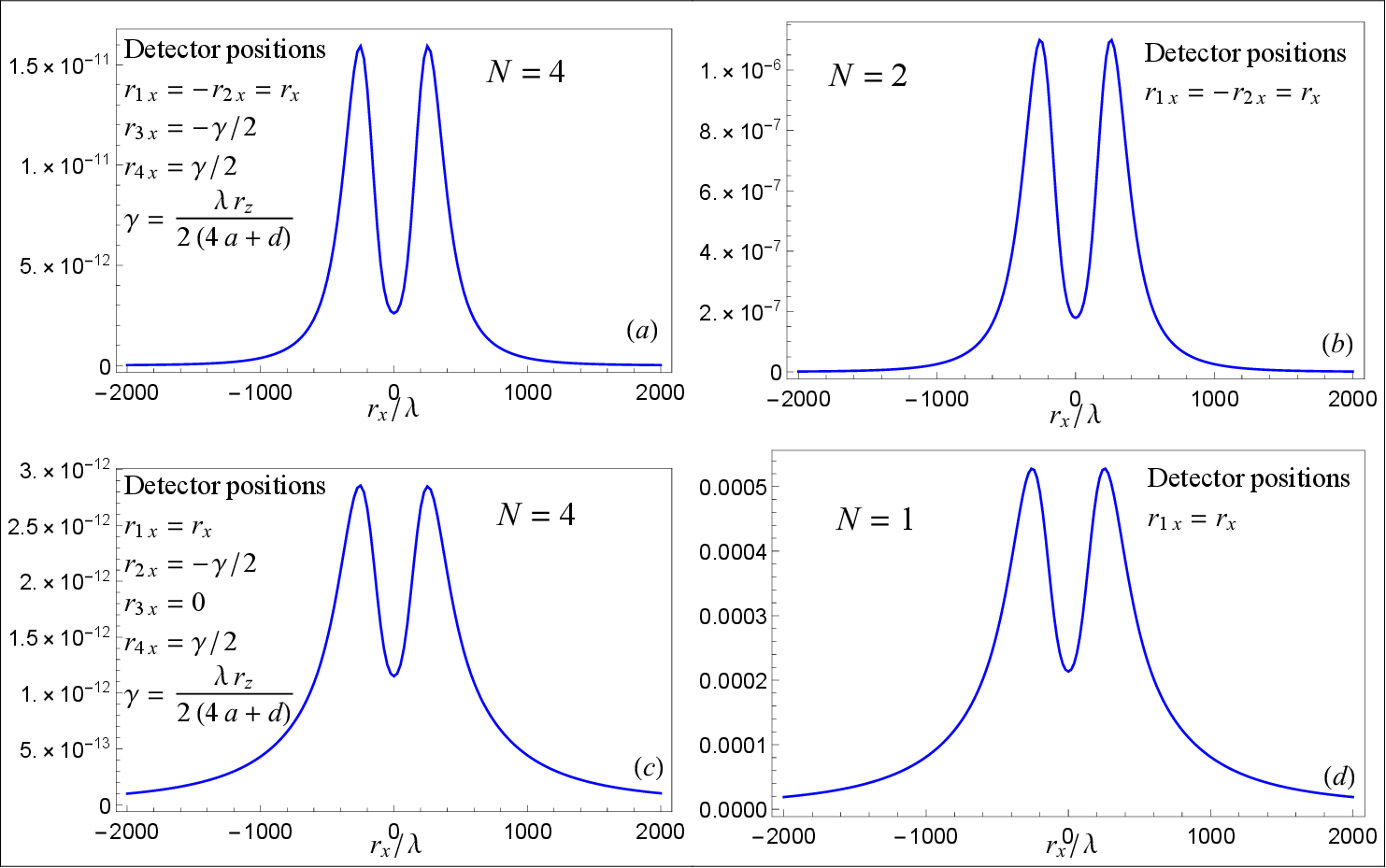}
\caption{$(a)$ shows the cross section of the superresolution pattern in the $x-$ direction for a four-emitter four-detector setup. In $(a)$, two detectors are in motion while the others remain stationary, which yields the same superresolution as a two-emitter two-detector configuration depicted in $(b)$. In $(c)$, the pattern is the same as a single emitter single detector configuration, as only one detector is in motion while the others stay still. The signals in $(a)$ and $(c)$ are $G^{(4)}(\mathbf{r})$, while $(b)$ and $(d)$ are $G^{(2)}(\mathbf{r})$ and $G^{(1)}(\mathbf{r})$ respectively. The strength of the signals is measured in arbitrary units. In all, $R_z=\lambda/10$ and $r_z=500\lambda$.
}
\label{sfig4}
\end{figure}

Figs. \ref{sfig4} $(a)$ and $(c)$ illustrate the superresolution patterns for a four-atom, four-correlated detector setup. In the former, two detectors are in motion, while the other two remain still. In the latter, one detector is in motion and the other three remain stationary.
Fig. \ref{sfig4} $(a)$ and $(c)$ are compared with $(b)$ and $(d)$ respectively. $(b)$ is the superresolution pattern for a two-atom, two-correlated detector configuration with two detectors in motion, while $(d)$ is one emitter with one detector. We mean coincidence detections by correlated detector measurements. The images in the same row have the same resolution as each other. The efficiency of multiatom and multidetector resolution will be contingent on the connections between the detectors and the manner in which they are operated. The way the correlated detectors operate in the same setup can be altered to manipulate quantum superresolution \cite{Thiel2007}. The results have demonstrated this.

\end{document}